# Remote atrial fibrillation burden estimation using deep recurrent neural network

Armand Chocron, Julien Oster, *IEEE member*, Shany Biton, Mandel Franck, Meyer Elbaz,
Yehoshua Y. Zeevi, and Joachim Behar, *IEEE member*

**Abstract— *Objective*:** The atrial fibrillation burden (AFB) is defined as the percentage of time spend in atrial fibrillation (AF) over a long enough monitoring period. Recent research has demonstrated the added prognosis value that becomes available by using the AFB as compared with the binary diagnosis. We evaluate, for the first time, the ability to estimate the AFB over long-term continuous recordings, using a deep recurrent neutral network (DRNN) approach. *Methods:* The models were developed and evaluated on a large database of p=2,891 patients, totaling t=68,800 hours of continuous electrocardiography (ECG) recordings acquired at the University of Virginia heart station. Specifically, 24h beat-to-beat time series were obtained from a single portable ECG channel. The network, denoted ArNet, was benchmarked against a gradient boosting (XGB) model, trained on 21 features including the coefficient of sample entropy (CosEn) and AFEvidence. Data were divided into training and test sets, while patients were stratified by the presence and severity of AF. The generalizations of ArNet and XGB were also evaluated on the independent test PhysioNet LTAF database. Results: The absolute AF burden estimation error $|E_{AF}(\%)|$, median and interquartile, on the test set, was 1.2 (0.1-6.7) for ArNet and 3.1 (0.0-11.7) for XGB for AF individuals. Generalization results on LTAF were consistent with $|E_{AF}(\%)|$ of 2.6 (1.1-14.7) for ArNet and 3.6 (1.0-16.7) for XGB. *Conclusion:* This research demonstrates the feasibility of AFB estimation from 24h beat-to-beat interval time series utilizing recent advances in DRNN. *Significance:* The novel data-driven approach enables robust remote diagnosis and phenotyping of AF.

## I. INTRODUCTION

Atrial fibrillation (AF) is the most common arrhythmia, with an estimated prevalence of 3% in adults aged 20 years or older [1], [2] and has an increased prevalence with aging [3]. AF is associated with quivering or irregular heartbeat, that can lead to blood clots, stroke, heart failure and other heart-related complications [4]. The estimated direct costs related to AF amount to 1% of the total healthcare spending in the UK and between 6-26 billion US dollars in the US for the year 2008 [5]. Existing treatments for AF include cardioversion and cardiac ablation as well as drugs intending at controlling the heart rate [6]. On the electrocardiogram (ECG) signal AF is characterized by an irregular heart rate and by the absence of p-wave. The currently accepted convention for AF diagnosis is the presence of an episode lasting at least 30 seconds on the ECG [6] and thus the diagnosis for AF is categorical (paroxysmal or persistent), meaning that eventually a patient is either classified as AF or non-AF. However, AF may be characterized with further details over longer continuous recordings, lasting from hours to days to better phenotype the condition. Long recordings should facilitate the analysis of the temporal distribution of AF events [7] or to compute the percentage of time spent in AF versus non-AF rhythm, a measure referred as AF burden (AFB) [8] [9]. In addition, many individuals with AF go undetected and thus are not treated because they are considered to be asymptomatic or to have paroxysmal AF (PAF) i.e. episodes of AF that occur occasionally. This motivates long term continuous recordings to be performed in order to enhance the diagnostic of AF. Estimation of the AFB may be achieved through the usage of wearables within the context of remote health monitoring [10] and the analysis of fiducials on the signals that are robust to noise and enabling to detect AF events; *Using wearables:* with the recent development of portable medical sensors, it became relatively easy to collect long term recordings of ECG (e.g. Zio patch (iRhythm technologies)) or photoplethysmography (e.g. SmartWatches) data. *Using a stable fiducial:* the signals recorded by portable sensors can be noisy or vary significantly with respect to the hardware manufacturer. We therefore decided to analyze the beat-to-beat interval time series because the R-peak is a robust fiducial that can be detected. We hypothesize that the AFB can be accurately estimated from the beat-to-beat time series, using a data-driven approach. The work presented herein contributes the following: (1) evaluating the feasibility, for the first time, of using deep recurrent neural network for accurately estimating the AFB from long continuous beat-to-beat interval time series; (2) providing a comprehensive and rigorous benchmark of our approach against state-of-the-art feature-engineering-based models; (3) producing, for the first time, comprehensive learning curves to understand the added value of growing training datasets in performing the task at hand; (4) evaluating the performance of the algorithm developed on an independent database to evaluate further its ability to generalize.

## II. MACHINE LEARNING FOR AF DETECTION

Various machine learning algorithms and approaches for physiological time series analysis have been previously experimented. These include classical machine learning such as support vector machine and random forest that use engineered features based on prior physiological knowledge and thus are more easily interpreted, to deep neural networks (DNN) approaches such as convolutional neural network (CNN). Deep learning models aim to perform within a single framework the feature extraction process from the raw data and the final classification process. CNNs, widely used in the field of computer vision, are able to leverage the characteristic visual patterns between adjacent data points, to build robust features for classification. CNNs may be used with 1-D convolutions directly applied to the time series or 2-D convolutions applied to the transformation of the time series into an 2D representation (image) [11]. The motivation behind the application of time series to image transformation approach is to harness the substantial advances made in the development and performance of CNNs in the field of computer vision. Recurrent neural networks (RNN) such as long short-term memory (LSTM) may also be used for the purpose of classification. RNN have the advantage of better capturing the time dependencies between consecutive windows that are being analyzed which bears meaning in many instances. Finally, hybrid methods combining classical machine learning, CNN and RNN architectures may be elaborated to take advantage of each individual approaches. A number of approaches drawing from these techniques have been experimented for physiological time series but it is unclear which of these are best both in terms of performance and interpretability. For example, the past four years PhysioNet/Computing in Cardiology (CinC) Challenges, a yearly worldwide competition on the topic of physiological time series analysis, have seen winning entries drawing from classical feature engineering-based classifiers [6], [7], deep learning [8], [9] and mixture of both approaches through ensemble learning [10], with no clear consensus on what performs best. This lack of clarify in ML models objective comparison has been highlighted in many sub-field of ML including deep metric learning [12] and adversarial trained deep networks [13]. In our research we offer an objective and rigorous comparison of different machine learning strategy.

For the 2017 CinC competition on the topic of AF detection from single lead ECG there were four winners. All made use of features engineered from the heart rate variability (HRV) and the morphology of the waveform [14]. In the context of AF diagnosis from HRV alone, a noticeable work includes single features drawing from information theory such as the coefficient of sample entropy (CosEn) [15] or engineered from the Lorenz Plot [16] used by Medtronic. The combination of multiple features within classical machine learning model has also been evaluated by Carrara et al. [17]. DNN have also been used: Faust et al. [18] developed a bidirectional network to assess the presence of AF among patients (p=25) based on the RR intervals, but the network capacity has not been verified on a large amount of data. Furthermore, their approach provides a binary output rather than an estimation of the AFB. Hannun et al. [19] developed a DNN to detect arrhythmias from ECGs including AF. While this approach performed well on a large amount of data (n=91,232, p=53,549), it was not designed and assessed to analyze long-term recordings which include long track of non-arrhythmic data. A comprehensive review on the usage of deep learning in ECG analysis, including AF detection, can be found in Hong et al. [21]. To date, and by opposition to the field of computer vision, it is unclear what algorithmic approach, e.g. classical feature-based learning versus DNN, is best both in term of performance and in term of their interpretability for the purpose of AF detection. Furthermore, the feasibility to estimate the AFB from beat-to-beat time series leveraging long-term and continuous recordings has not been researched.

## III. METHODS

### A. Databases

*Model development database*

The University of Virginia Database (UVAF) [22], [23] consists of RR interval time-series and rhythm annotations. This database gathers the electrocardiogram (ECG) recordings of patients over 39 years ($46.4 \pm 25.5$ years) for whom University of Virginia health system physicians ordered Holter monitoring from December 2004 to October 2010. Indications for the Holter recordings included palpitations (40%) or syncope and dizziness (12%). This database contains p=2,891 annotated files of individual patients. Each file contains relative timestamps for each heartbeat, their type and rhythm type. These annotations were automatically generated by the medical monitor Philips Holter software, based on the automated analysis of the ECG trace. Part of these files (52.3%) were reviewed by medical school students. Each record lasts approximately 24 hours ($23.7 \pm 1.75$ hours). The distributions in Fig. 1 summarize the frequency of the rhythm's annotations associated with each RR interval. This is shown for the most represented abnormal rhythms categories: AF, sinus bradycardia (SBR), supraventricular tachyarrhythmia (SVTA), ventricular bigeminy (B) and ventricular trigeminy (T). AF is well represented which ensures that this database is suitable to evaluate the AFB from long RR time series. We considered atrial flutter (AFL) to be the same class as AF similar to Carrara et al. [17]. Recordings were divided in 60-RR windows for analysis. This number (60-RR) is justified by the medical diagnosis criteria which states that a patient is diagnosed as positive AF for an episode lasting at least 30 seconds [6] and assuming a beating rate in the range 60-120 bpm this corresponds to up to 60 RR intervals.

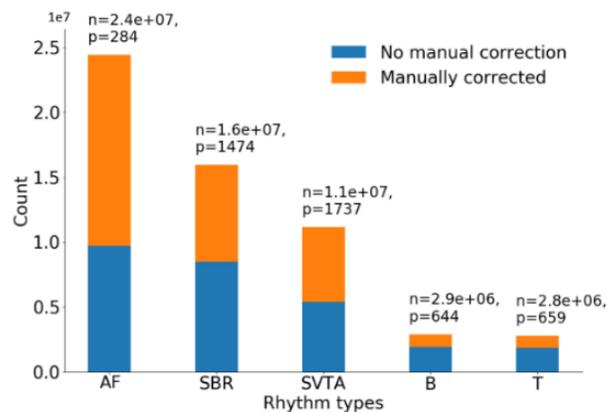

**Fig. 1.** Distribution of the rhythm annotations for each individual beat across the UVAF dataset. The most represented rhythms annotations are shown. In blue, the beat annotations which have not been manually verified; in orange, the beat annotations that were manually verified. The normal (N) beats (n=2.7e+08, p=2805) are not displayed for visual purpose.

*Independent database for generalization*

The PhysioNet Long-Term AF (LTAF) [24] database is used as an independent test set to evaluate the generalization of the models. LTAF consists of recordings of p=84 individuals suffering from PAF or sustained AF. Each record contains two simultaneously recorded ECG signals digitized at 128 Hz, with 12-bit resolution over a 20 mV range; record durations are of $22.7 \pm 2.4$ hours long. The overall database totals t=1,900 hours, including 874 hours in AF and 1,026 hours spent in non-AF rhythms. The original recordings were digitized, automatically annotated at Boston's Beth Israel Deaconess Medical Center, Boston, and further manually reviewed.

## B. Classes definition

The currently accepted convention for AF diagnosis is the presence of an episode lasting at least 30 seconds [6]. Boriani et al. [25] found that maximum daily AF burden longer than 1h carries important negative prognostic implications and may serve as a clinically relevant parameter and improve the assessment of a risk stratification for stroke. This means that individuals with an AFB greater than 4% will have increased prognostic of having a stroke. Based on these research and guidelines, we defined for each patient an overall rhythm label defined as follows: a patient was labelled as mild PAF ($AF_{mild}$) if his AFB was in the range ≥30 seconds and up to 4% of the time over the whole recording length. A patient was labelled as moderate PAF ($AF_{mod}$) if his AFB was in the range 4-80% over the whole recording length. A patient was labelled as severe AF ($AF_{sev}$) if his AFB was over 80% of the whole recording time. Individuals who had less than 30 seconds of AF were labelled as non-AF. The number of individuals falling in each of the defined groups is summarized in TABLE I. This differentiation was made to best stratify the training and test sets individuals and in order to evaluate the algorithms for the different groups (non-AF, $AF_{mild}$, $AF_{mod}$ and $AF_{sev}$). Fig. 2 shows the distribution of the percentage of time spent in AF for the different groups studied, while Fig. S2 presents the distribution of the events over the UVAF database. These figures highlight that AF is not a binary diagnosis but rather that there exists a spectrum of AF levels or phenotypes.

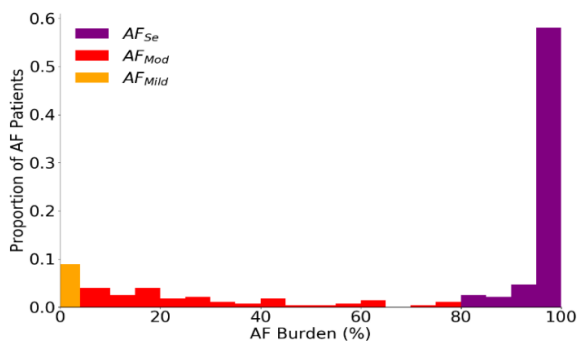

**Fig. 2.** Normalized histograms of AF burden for each AF group. This figure highlights the diversity of AF phenotypes in term of AF burden for the paroxysmal AF individuals.

TABLE I.
DESCRIPTION OF THE UVAF PATIENTS ACCORDING TO THE PRESENCE OF AF AND SEVERITY LEVEL.

|  | Patients (n) | Windows (p) |
|---|---|---|
| **Non-AF** | 2,612 (90.3%) | 5,007,501 (90.3%) |
| **Mild AF** | 25 (0.8%) | 42,866 (0.8%) |
| **Moderate AF** | 62 (2.1%) | 116,500 (2.1%) |
| **Severe AF** | 192 (6.7%) | 375,216 (6.7%) |
| **Total** | 2,891 | 5,542,083 |

## C. Data quality improvement

The raw RR time series, based on the annotations collected by the Philips Holter Software, was divided into non-overlapping windows of 60-RR. A total of 5,542,083 windows were available from the UVAF. For each individual window, we defined its rhythm label as the reference label most represented over this window. In order to automatically assess the quality of the raw ECG files and discard those that were too noisy, we adopted the R-peak quality criterion $bSQI$ [26]. For each window, the $bSQI$ index was computed with an agreement window of 50 $ms$. Briefly, the $bSQI$ index compares the R-peaks detected by two different R-peak detectors: one reference set, usually coming from a stronger R-peak detector, and one second set, coming from another, and usually weaker R-peak detector. If the two detectors agree (detect the same beats), then the quality can be assumed to be sufficiently high to reliably use the beat-to-beat time series. We used the $epltd$ implementation of the Pan and Tompkins algorithm [27] for reference R-peak annotation and $xqrs$ [24] for test R-peak annotation. We verified that the generated annotations contained at least 1,000 R-peak and excluded files which did not satisfy this criterion i.e. typically corresponding to recordings with a flat ECG. Furthermore, some reference annotations were missing over some windows. In order to account for that, windows with over 10 seconds of missing reference annotations were excluded. Patients presenting over 25% of missing windows based on this criterion were excluded from the analysis. Among the remaining patient recordings, windows with a $bSQI$ lower than 0.8 were excluded from the analysis [26]. Recordings showing a rate of exclusion, i.e., the ratio between the number of excluded windows and the total number of windows, higher than 75% were considered as corrupted and were discarded. Fig. 3 summarizes the steps adopted for handling missing data and excluding low quality signals.

## D. Machine learning for AF events detection

### Training and test sets

The UVAF dataset was divided into training and test sets with a ratio of 80-20%. To ensure the quality of the annotations of the UVAF test set (Fig. 3) we selected for the test set only patients whose ECG annotations were reviewed. The patients were stratified according to their overall rhythm label and the severity of AF (see section II.B). One of the key aspects to consider while evaluating a ML model is its ability to generalize on independent datasets that are derived from different population samples as well as data recorded with different devices. For that purpose, the models were further evaluated on the LTAF [24] independent test dataset.

### Feature selection

We identified 21 features (TABLE II) which have been widely used in the field of AF detection from RR time series, among them: Coefficient of Sample Entropy (CosEn) [15], AFEvidence (AFEv) [16], Origin Count in the Lorenz Plot (OC) [16], Irregularity evidence (IrrEv) [16] and Premature Atrial Complex Evidence (PACEv).

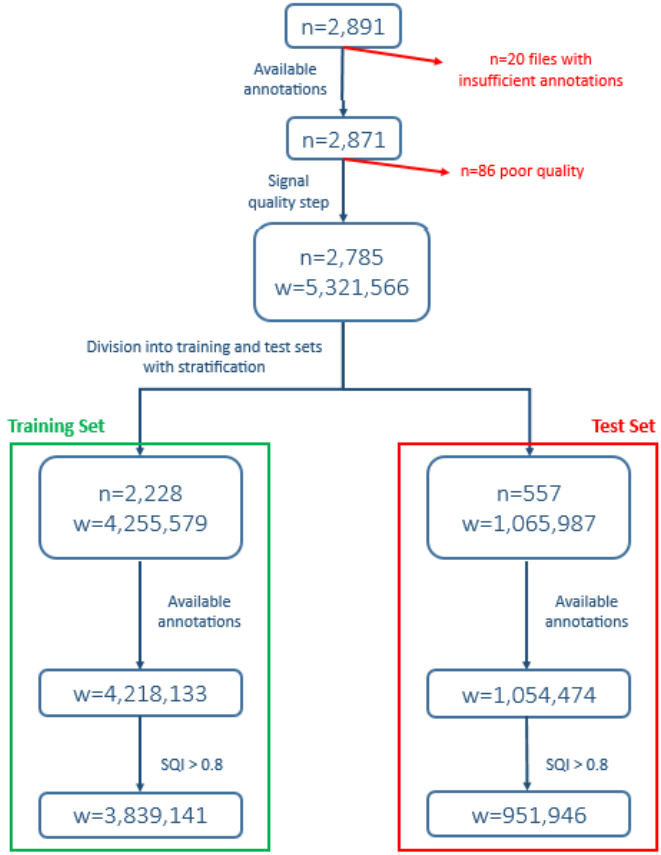

**Fig. 3.** Data exclusion and stratification process applied to the UVAF database.

| Number | Feature | Definition |
|--------|---------|------------|
| 1 | CosEn | Coefficient of sample entropy [15] |
| 2 - 5 | AFEv, IrrEv, PACEv, OriginCount | Measures derived from the Lorenz plot to assess irregularities in the RR intervals. [16] |
| 6 - 7 | PoincSD1, PoincSD2 | The standard deviation on the two principal axis of the ellipse on the Poincare plot. [29] |
| 8 | minRR | The minimal RR interval in the segment. |
| 9 | medHR | The median heart rate in the segment. |
| 10 | AVNN | The mean RR interval over the segment. |
| 11 | SDNN | The standard deviation of the RR intervals over the segment. |
| 12 | SEM | Standard error of the mean. |
| 13 - 14 | PNN20, PNN50 | The percentage of RR intervals shorter than 20 and 50 [ms], respectively. [30] |
| 15 | RMSSD | The root mean square of the successive differences. |
| 16 | CV | Coefficient of variation. |
| 17 - 20 | PIP, IALS, PSS, PAS | Fragmentation measures. [31] |
| 21 | bSQI | The signal quality index over the window. [26] |

The CosEn is computed as follows:

$$CosEn = -\ln\left(\frac{A}{B}\right) - \ln(2r) - \ln(\overline{RR}), \quad (1)$$

where $A$ and $B$ represent the number of matching segments of sizes m+1 and m respectively ($m$=2), $r$ is the matching tolerance interval ($r$=0.03 [s]), $\overline{RR}$ is the mean RR interval over the segment [15]. The AFEv feature is computed as follows [16]:

$$AFEv = \sum_{n=1}^{12} BC_n - OriginCount - 2 \times PACEv, \quad (2)$$

where $BC_n$ is the number of points in the $n^{\text{th}}$ segment of the Lorenz plot, OriginCount is the number of points in the center bin and, finally PACEv is an additional measure representative of the ectopic beats patterns [16]. Features were standardized with regard to the mean and the standard deviation, as estimated from the training set of examples. We accounted for the imbalance between the classes by means of weights that were inversely proportional to their representation in the training set. Several classification algorithms were benchmarked: threshold based classification for individual features, logistic regression (LR), random forests (RF) using the python scikit-learn library and gradient boosting (XGB) [28]. A binary classification of AF examples (consisting of windows of 60-RR) was considered. The regularization coefficient ($C$) for LR as well as the maximal depth of the trees ($m_d$) and the number of estimators ($n_e$) for RF and XGB were optimized using 5-fold cross-validation on the training set (Table SI).

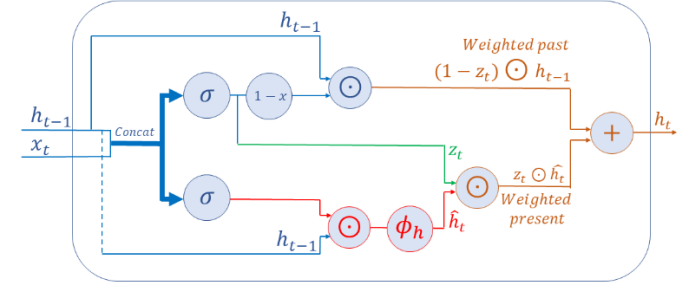

**Fig. 4.** A simplified schematic diagram of a Gated Recurrent Unit (GRU). The previous state $h_{t-1}$ and the current input $x_t$ are fed into input and reset gates, that achieve the proper balance between present and past information. Previous state $h_{t-1}$ and the estimated current state $\hat{h}_t$ are eventually weighted to provide the final input of the cell.

*Deep learning*
Recurrent Neural Networks (RNNs) and, in particular, Gated Recurrent Units (GRUs) are capable of dealing with temporal data using built-in memory units, which accurately select relevant past information (Fig. 4). In order to benefit from the advantages of both CNN and GRU networks, a model combining both architectures has been designed. We first trained a 1D-CNN model on the individual 60-RR windows, independent of their temporal relationship. This 1D-CNN included 2 convolutional layers followed by a local max pooling of size $1 \times 2$, followed by an additional convolutional layer. All the filters had a fixed size of $1 \times 10$. These CNN layers were followed by 3 fully connected layers. The number of filters $n_{filt}$ used in the first layer, as well as the number of hidden units $n_{hid}$ in the first fully-connected layer were selected via 5-fold cross validation (TABLE SI). The number of filters from a layer to the subsequent one was gradually increased by a factor two, while the number of neurons from a fully-connected layer to another was divided by two. A flowchart of the 1D-CNN is shown in Fig. 5. Once the 1D-CNN trained, we fed the $n_{hid}$ features generated at the first fully-connected layer to a GRU. Depending on the AFB estimated by the 1D-CNN, a GRU is selected from a pool of four GRU models: $GRU_{Non-AF}$, $GRU_{Mild}$, $GRU_{Mod}$, and $GRU_{Sev}$. The insight being that each GRU better captures the temporal distribution of AF events for a given AF severity level, based on its own training examples. Generally speaking, a GRU layer processes the input data from $t = -(h-1), ..., 0$ where $h$ is defined as the history length. The output of the layer is produced as follows:

$$z_t = \sigma(W_z x_t + U_z h_{t-1} + b_z) \qquad (3)$$

$$r_t = \sigma(W_r x_t + U_r h_{t-1} + b_r) \qquad (4)$$

$$\widehat{h_t} = \phi_h(W_h x_t + U_h(r_t \odot h_{t-1}) + b_h) \qquad (5)$$

$$h_t = (1 - z_t) \odot h_{t-1} + z_t \odot \widehat{h_t} \qquad (6)$$

The input vector $x_t$ is fed into both a gate and a reset layer whose outputs are $z_t$ and $r_t$ respectively. These outputs define a fine-tuned balance between the current and the previous states, to produce the final output $h_t$. $\sigma$ and $\phi_h$ represent sigmoidal and hyperbolic tangent functions which play the role of activation functions, and $\odot$ denotes the Hadamard product. The length of the output $h_t$ ($n_{units}$) was selected using 5-fold cross-validation. The history length $h$ has been searched using 5-fold cross-validation. Each one of the different GRU units has the same architecture (Fig. 4). At the training stage, these different units are fed with the windows belonging to the target population ($Non-AF, AF_{Mild}, AF_{Mod}, AF_{Sev}$). At the inference stage, a given patient is assigned to a population in accordance with the result of the AFB evaluation, based on the 1D-CNN. The corresponding windows are then fed to the selected GRU model, and this based on the AFB estimate (Fig. 5). We term the corresponding trained model "ArNet". The extended deep learning architecture, incorporating GRUs, was trained on an NVidia GeForce RTX 2080.

### AF burden estimation

AF burden is defined as the percentage of time an individual is in the state of AF during a long-enough monitoring period, i.e.:

$$AFB = \frac{\sum_{i=1}^{N} l_i \times \mathbb{1}_i}{\sum_{i=1}^{N} l_i}, \qquad (7)$$

where $l_i$ is the length of the $i^{th}$ window in seconds, $N$ is the number of available windows, and $\mathbb{1}_i$ is the unity operator which is equal to 1 when the window is AF and zero otherwise.

### Performance statistics

In order to assess the performance of the classifiers in classification of individual 60-RR examples, the following characteristics are computed: sensitivity ($Se$), specificity ($Sp$), positive predictive value ($PPV$), accuracy ($Ac$) and the harmonic mean between the $Se$ and $PPV$ termed $F_1$-Score. The classifiers are optimized to maximize the area under the receiver operating curve (AUC). The probability decision threshold is chosen as the point which maximizes the $F_1$-Score.

To evaluate the AFB accuracy, we define the error in AFB measure ($E_{AF}$) as:

$$E_{AF}(\%) = \frac{\sum_{i=1}^{N} l_i \times (\hat{y}_i - y_i)}{\sum_{i=1}^{N} l_i} \qquad (8),$$

where $y_i$ represents the reference label of the window (1 for AF, 0 otherwise), $\hat{y}_i$ represents the label predicted by the model (1 for AF, 0 otherwise). This error was computed for each group (non-AF, $AF_{mild}$, $AF_{mod}$ and $AF_{sev}$). For each group the 5-summary statistics of $|E_{AF}|$ are reported.

### Learning curves

Learning curves were used in order to analyze how the ML algorithms perform with a growing number of examples. The increase in the number of examples was done sequentially by adding patients to the training set. This routine was applied so that the learning curves depicted the model performance as a function of the number of patients, and increasing variability from a more diverse population. Practically, we selected the windows used in training, while selecting them from an increasing subset of patients ranging from 20 to 2228. For a number of patients up to 558, i.e. twice the number of AF individuals, half of the patients are sampled among the AF patients and half among the non-AF individuals. When the number of AF cases was exhausted then only non-AF patients were added until the completion of the recordings from the training set.

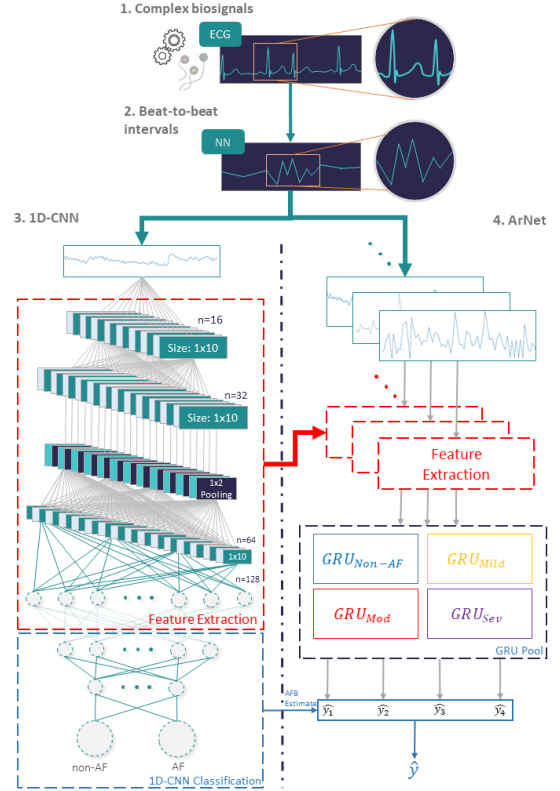

**Fig. 5.** Architectural layout of the CNN-based model for AF classification, termed "ArNet". Beat-to-beat interval time series are extracted based on data collected from a single channel portable electrocardiography device. These intervals constitute the inputs of a 1D-CNN, trained to classify AF events (left panel). The 1D-CNN is used as a feature extractor to build the features of h=10 consecutive windows, further fed to a Gated Recurrent Unit (GRU) pool containing 4 units, whose outputs constitute candidates for the final prediction The output is selected based on the 1D-CNN AFB estimate for the given patient.

## IV. RESULTS

*Data preprocessing and exploration*

Among the 2,891 original files, p=20 patients were discarded because the percentage of missing annotations exceeded 25% (Fig. 3). An additional p=86 patients presenting over 75% of their windows corrupted were discarded. These two steps resulted in the exclusion of 106 patients i.e. 3.6% of the total number of patients in UVAF. The remaining patients were stratified with respect to their sub-group i.e. Non-AF, $AF_{Mild}$, $AF_{Mod}$, $AF_{Se}$. For each recording, the windows having annotations available and a $bSQI > 0.8$ were kept. This resulted in a total of $w = 4,791,087$ windows being kept i.e. 90.0% of all available windows. Features were computed on these windows. The distribution of the most discriminative engineered features for the different groups are displayed in Fig. S1 and provides some insights about the discriminative power of these features in separating AF from non-AF examples.

*Models performance*

The optimal hyperparameters obtained after 5-fold cross validation for the 1D-CNN and ArNet were $n_{filt} = 64$, $n_{hid} = 128$, $h = 10$. The results regarding the classification of 60-RR windows as AF or non-AF during the cross validation and for the test set are summarized in Table SII and TABLE III. The best test score was obtained for ArNet with $F_1 = 0.92$ versus $F_1 = 0.88$ for the best feature-based engineering model (XGB). The AF burden estimation error ($E_{AF}$) median and interquartile (Q1-Q3), on the test set, was 1.2 (0.1-6.7) for ArNet and 3.1 (0.0-11.7) for XGB in estimating the per patient AFB for AF individuals. TABLE IV presents the detailed statistics regarding the AFB estimation error and Fig. 6 the histogram of the error for each one of the different groups. Generalization results on LTAF were consistent with $E_{AF}$ of 2.6 (1.1-14.7) for ArNet and 3.6 (1.0-16.7) for XGB.

TABLE III.
PERFORMANCE STATISTICS FOR SINGLE WINDOW CLASSIFICATION IN UVAF TRAIN AND TEST SETS (SEE FIG. 3).

|  |  | $F_1$ | AUC | Se | Sp | PPV |
|---|---|---|---|---|---|---|
| **TRAIN** | **LR** | 0.77 | 0.97 | 0.78 | 0.98 | 0.77 |
|  | **RF** | 0.8 | 0.98 | 0.79 | 0.99 | 0.82 |
|  | **XGB** | 0.83 | 0.99 | 0.82 | 0.99 | 0.85 |
|  | **1D-CNN** | 0.87 | 0.99 | 0.87 | 0.99 | 0.88 |
|  | **ArNet** | 0.89 | 0.99 | 0.89 | 0.99 | 0.89 |
| **TEST** | **LR** | 0.83 | 0.97 | 0.8 | 0.99 | 0.85 |
|  | **RF** | 0.85 | 0.98 | 0.82 | 0.99 | 0.89 |
|  | **XGB** | 0.88 | 0.99 | 0.85 | 0.99 | 0.91 |
|  | **1D-CNN** | 0.89 | 0.99 | 0.88 | 0.99 | 0.9 |
|  | **ArNet** | **0.91** | **0.99** | **0.90** | **0.99** | **0.91** |

TABLE IV.
ABSOLUTE AF BURDEN ESTIMATION ERROR ($|E_{AF}|$ [%]) FOR THE UVAF TEST SET AF PATIENTS (P=66).

|  | Min | Q1 | Med | Q3 | Max |
|---|---|---|---|---|---|
| **LR** | 0.1 | 2.7 | 6.4 | 23.1 | 96.8 |
| **RF** | 0.01 | 0.8 | 4.1 | 20.7 | 99.7 |
| **XGB** | **0.0** | 0.0 | 3.1 | 11.7 | 96.8 |
| **1D-CNN** | 0.01 | 0.9 | 2.2 | 12.1 | 96.8 |
| **ArNet** | **0.0** | **0.1** | **1.2** | **6.7** | 98.8 |

## V. DISCUSSION

Our first major conclusion is that it is possible to accurately estimate the AFB using long term beat-to-beat time series analysis that may be collected remotely from a single ECG channel. The best results in terms of $E_{AF}$ were obtained by ArNet, with a median of 1.2 (0.1-6.7) for the AF cases on the UVAF test set. Results were consistent on the LTAF independent test dataset. Our second major conclusion is that the deep recurrent neural network, ArNet, provides the best results in estimating the AFB versus feature-based machine learning models. To benchmark ArNet, we reviewed and implemented HRV features that have proven to be discriminative of AF rhythm [15], [16], [31]. We therefore integrated them into state-of-the art machine learning models (RF, XGB) to benchmark against DNN. This was the case in both classification of individual 60-RR windows (TABLE III) as AF or non-AF, and in evaluation the AFB (TABLE IV). ArNet outperforms the 1D-CNN using the temporality between windows ($h = 10$), showing there is value in keeping historical information. Incidentally, this research establishes that recent advances in deep learning outperform state-of-the-art work developed over the past 20 years in HRV feature engineering for the task of AF classification from RR time series. The third major conclusion is that increasing the number of patients in the training set improved the performance of all models importantly up to 1200 patients (Fig. 6). The gain between 1200 and 2228 patients was moderate for ArNet (+0.05), XGB (+0.05) and RF (+0.03) but none for LR. It is, however, important to note that the number of AF patients in the training set being limited to 279, this means that from p=558 (dotted vertical line in Fig. 7). AF patients were not added anymore to the training set and only non-AF patients were added. It can be expected that adding more AF individuals to the training set will increase further the performance of the model. Another important observation based on the learning curves is that ArNet outperforms the benchmark models for more than p=100 individuals in the training set. This is in accordance with findings in the field of machine learning per which deep neural network outperform classical machine learning models provided enough data are available, while feature engineering-based algorithms are more adapted to classification problems for which only little data is available.

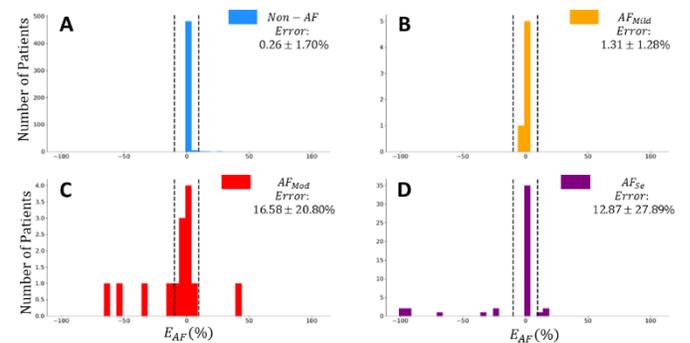

**Fig. 6:** AF burden estimation error of ArNet for the different classes of individuals: Non-AF (panel A), $AF_{Mild}$ (panel B), $AF_{Mod}$ (panel C), and $AF_{Se}$ (panel D). The dotted black lines represent the overall standard deviation for the respective class

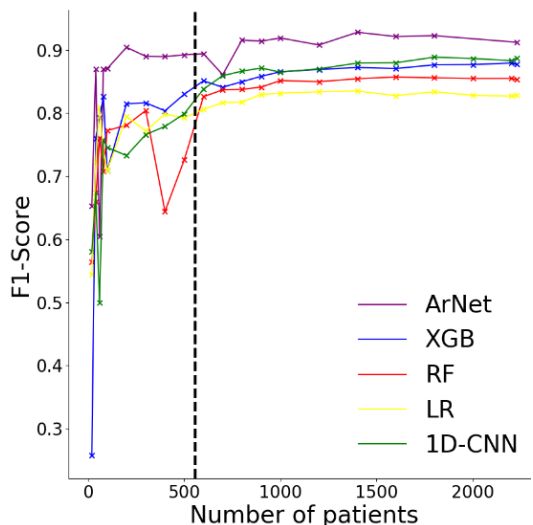

**Fig. 7.** Learning curves of the different algorithms which represents the performance ($F_1$-Score) of the algorithm on the test set with regard to the training set size. The vertical black dashed line represents the point at which all training set AF patients have been added and thus after this line only non-AF individuals are added to the training set.

### Impact on binary classification

The assessment of the AFB based on long-term recordings enables the identification of additional cases which would be missed by a standard 12-lead ECG. Indeed, patients which suffer from PAF are likely to be missed as they present only sporadic and short events across the day. For example, setting a hard threshold at 4% on the predicted AFB for the purpose of binary classification, we report the following performance measures on the test set: $F_1 = 0.84$, $Se = 0.85$, $Sp = 0.98$. In comparison, using the prediction of ArNet on a randomly selected window, we report $F_1 = 0.78$, $Se = 0.66$, $Sp = 0.99$. A total of 11/66 additional cases (17%) were identified using the AFB approach for performing a binary diagnosis. This further stress the importance of leveraging long-term recordings to perform a robust AF diagnosis.

### Error analysis

As an attempt to understand the limitations of the algorithm in its capacity to correctly estimate the AFB, we represented in Fig. 8 the FP detected windows of 60-RR per rhythm type in an attempt to identify which rhythms are often misclassified as AF. In particular, this analysis revealed that SVTA accounts for a significant number of the FP with up to 3.4% of the SVTA cases that were misclassified as AF. This can be explained by the fact that AF often manifests as an abnormally high heart rhythm. Besides, the mean proportion of ectopic beats in the false positive cases (3.1%) was significantly higher compared to the false negatives (1.7%, p<0.01). This is in accordance with previous studies [32] which showed the that ectopic beats are often misclassified as AF. In addition, cardiologist manual review (co-author FM) of the five cases with the highest $|E_{AF}(\%)|$ revealed that 4 cases out of 5 presented a highly regular flutter while the remaining recording showed a slow AF rhythm with a median beating rate at 54 (±9) bpm. These cases point out an intrinsic limitation of ArNet whose output is based exclusively on the RR interval time series which might not be enough to robustly identify AFL.

### Limitations

Our approach has a number of limitations. First, 52.3% of the examples and rhythm annotations have been generated automatically by the Phillips Holter software and have not been reviewed properly by a medical expert. Further, although we demonstrated the generalization of our algorithm on the LTAF, these recordings were selected from larger databases and thus carry an intrinsic bias. Consequently, the first important limitation is the need for an external validation on an independent test dataset with a representative population sample. It has been shown that the generalization of machine learning algorithms is one of the main limiting reasons for their implementation in clinical practice. We intend to investigate this question in future works. The second important limitation of our approach is that individuals presenting atrial flutter will likely be missed by the model (refer to error analysis subsection). This is an intrinsic limitation of the approach which may be alleviated by adding morphological information captured from the underlying pulsatile raw signal. In addition, the mean of $|E_{AF}(\%)|$ for males (p=267) and females (p=290) patients were of 2.5 and 0.9 respectively, while the mean across elderly (over 60 years old, p=206) and young (under 60 years old, p=351) AF patients were of 1.0 and 2.7 respectively. The $|E_{AF}(\%)|$ metric for females (p<0.01) and young subjects (p<0.01) was significantly lower, showing that ArNet performance is sensitive to the target population. Finally, the high performances reached by ArNet may be moderated by the lack of interpretability of deep learning models versus feature engineering-based models. We argue that although this rational could be true for waveforms analysis (e.g. ECG), it is less impactful for HRV since most of the engineered HRV features are, even to date, hardly associated with a clear physiological function (e.g. of such debate [33]) and thus provide limited to no added value over a deep learning based approach within our specific context.

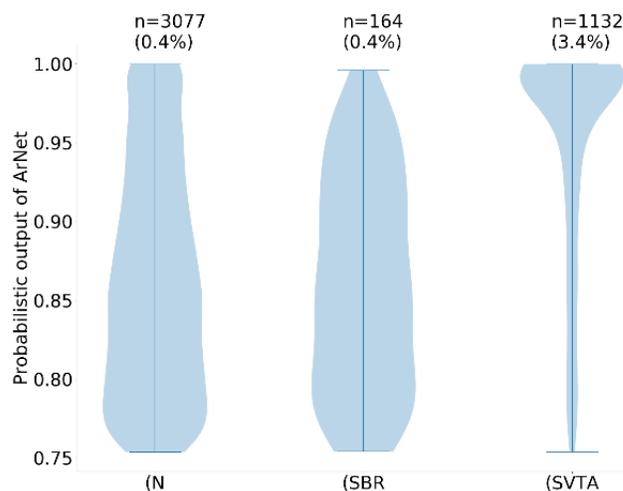

**Fig. 8.** Violin plots showing the probabilistic output of ArNet for the false positive examples (60-RR). The violin plots are shown for the different represented rhythms: Non-AF (N), Sinus Bradycardia (SBR), Supraventricular Tachycardia (SVTA). In particular, we observe that examples with SVTA rhythm are often misclassified as AF. Bigeminy (B) and Trigemiy (T) examples only had 7 and 8 FP respectively.

**Supplementary material**

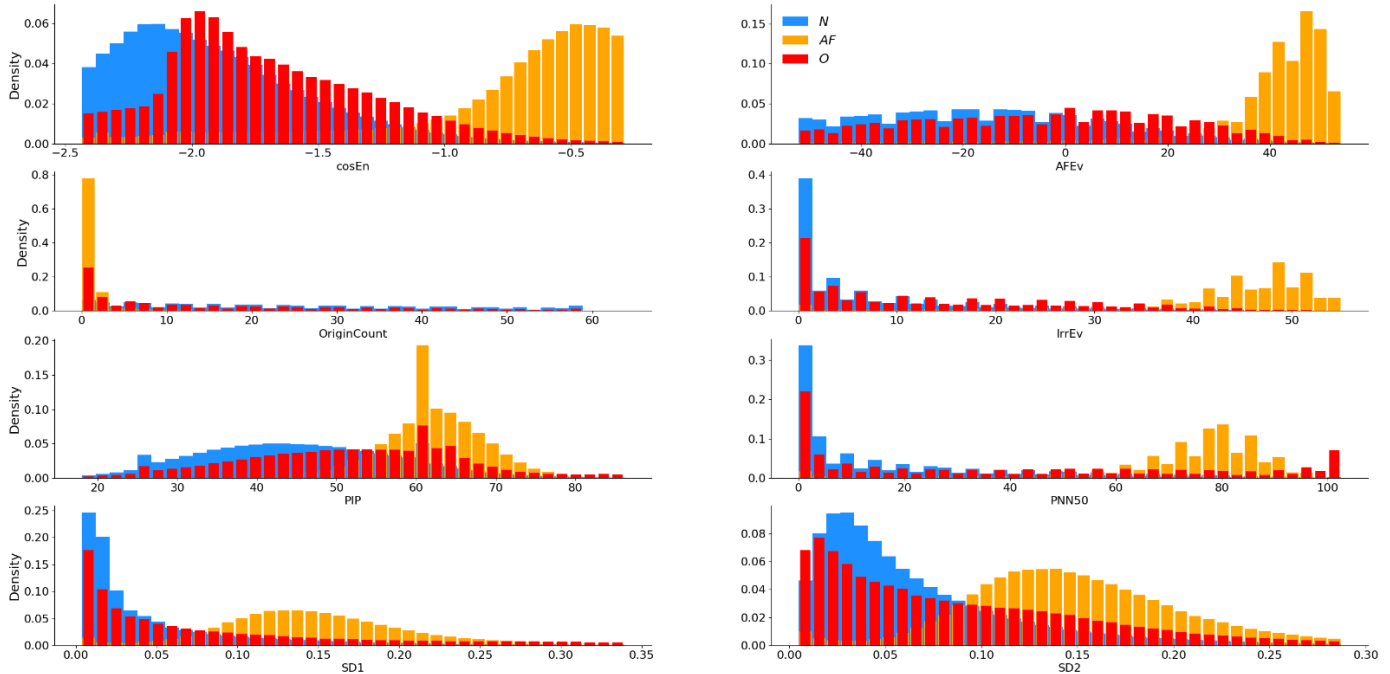

Fig. S1. Distributions of the most discriminative hand-crafted features further used by the different classifiers: cosEn, AFEv, OriginCount, IrrEv, PIP, PNN50, SD1 and SD2.

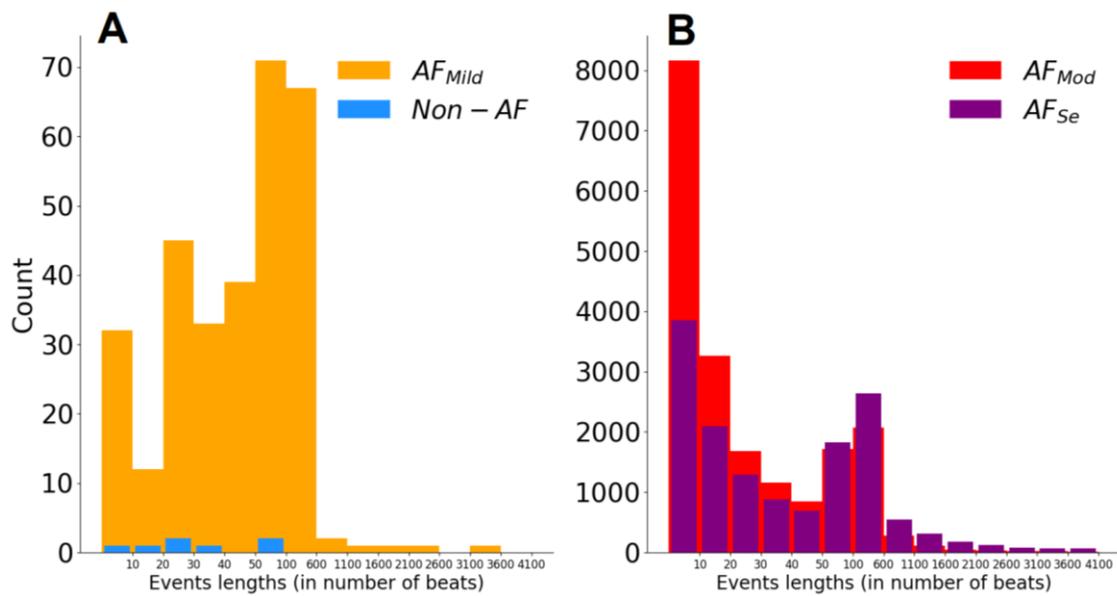

Fig. S2. Distribution of the events lengths among the different populations. The $AF_{pe}$ population has longer events as expected. The histogram shows the data up to the $98^{th}$ percentile, for visual purposes. The binning is the same for each one of the populations. There are overall n=34,971 events. A great number of events (n=24,861) present lengths smaller than 60 seconds (71.0%).

TABLE SI.
HYPERPARAMETERS GRID SEARCH FOR DIFFERENT MACHINE LEARNING MODELS.

|  | Hyperparameter | Search range | Search pace |
|---|---|---|---|
| **LR** | $C$ | 0.1 - 1000 | 10 - Exponential |
| **RF** | $m_d$ | 3 - 6 | 1 - Linear |
|  | $n_e$ | 70 - 120 | 10 - Linear |
| **XGB** | $m_d$ | 3 - 6 | 1 - Linear |
|  | $n_e$ | 70 - 120 | 10 - Linear |
| **1D-CNN** | $n_{filt}$ | 64 - 128 | 2- Exponential |
|  | $n_{hid}$ | 128 - 256 | 2 - Exponential |
| **ArNet** | $h$ | $5 - 20$ | 2 - Exponential |
|  | $n_{units}$ | 8 - 16 | 2 - Exponential |

TABLE SII.
RESULTS OF THE CROSS-VALIDATION FOR EACH ONE OF THE DIFFERENT ALGORITHMS.

|  | **F$_1$** | *AUROC* | *Se* | *Sp* | *PPV* |
|---|---|---|---|---|---|
| *LR* | 0.77 | 0.97 | 0.77 | 0.98 | 0.76 |
|  | (+/- 0.04) | (+/- 0.01) | (+/- 0.07) | (+/- 0.00) | (+/- 0.03) |
| *RF* | 0.80 | 0.97 | 0.78 | 0.99 | 0.82 |
|  | (+/- 0.05) | (+/- 0.01) | (+/- 0.09) | (+/- 0.00) | (+/- 0.03) |
| *XGB* | 0.82 | 0.98 | 0.81 | 0.99 | 0.83 |
|  | (+/- 0.04) | (+/- 0.01) | (+/- 0.07) | (+/- 0.00) | (+/- 0.03) |
| **1D-CNN** | 0.83 | 0.98 | 0.83 | 0.99 | 0.83 |
|  | (+/- 0.03) | (+/- 0.00) | (+/- 0.05) | (+/- 0.00) | (+/- 0.04) |
| **ArNet** | 0.90 | 0.99 | 0.88 | 0.99 | 0.92 |
|  | (+/- 0.03) | (+/- 0.00) | (+/- 0.05) | (+/- 0.00) | (+/- 0.04) |